\documentclass[twocolumn,showpacs,preprintnumbers,nofootinbib,aps]{revtex4}
\usepackage{graphicx,bm,latexsym,amssymb} 
\usepackage{amsmath,epsf,psfig}

\begin{document}

\newcommand{\hh}{\hspace{3mm}}

\newcommand{\sign}{\mbox{sign}}
\newcommand{\ssign}{\mbox{\scriptsize sign}}
\newcommand{\smin}{{\mbox{\scriptsize min}}}
\newcommand{\smax}{{\mbox{\scriptsize max}}}
\newcommand{\sch}{{\mbox{\scriptsize ch}}}
\newcommand{\tmin}{{\mbox{\tiny min}}}

\newcommand{\la}{\left\langle}
\newcommand{\ra}{\right\rangle}

\newcommand{\dla}{\la \!   \la}
\newcommand{\dra}{\ra \!   \ra}

\newcommand{\prtl}{\partial}
\newcommand{\we}{\widetilde}

\newcommand{\smfp}{{\mbox{\scriptsize mfp}}}
\newcommand{\smol}{{\mbox{\scriptsize mol}}}
\newcommand{\sph}{{\mbox{\scriptsize ph}}}
\newcommand{\sinhom}{{\mbox{\scriptsize inhom}}}
\newcommand{\sneigh}{{\mbox{\scriptsize neigh}}}
\newcommand{\srlxn}{{\mbox{\scriptsize rlxn}}}
\newcommand{\svibr}{{\mbox{\scriptsize vibr}}}
\newcommand{\smicro}{{\mbox{\scriptsize micro}}}
\newcommand{\selast}{{\mbox{\scriptsize elast}}}
\newcommand{\select}{{\mbox{\scriptsize elect}}}
\newcommand{\seq}{{\mbox{\scriptsize eq}}}
\newcommand{\scr}{{\mbox{\scriptsize cr}}}
\newcommand{\sT}{{\mbox{\scriptsize T}}}
\newcommand{\sTLS}{{\mbox{\scriptsize TLS}}}
\newcommand{\sd}{{\mbox{\scriptsize d}}}
\newcommand{\sext}{{\mbox{\scriptsize ext}}}
\newcommand{\scav}{{\mbox{\scriptsize cav}}}
\newcommand{\bmu}{\bm \mu}
\newcommand{\bE}{\bm E}
\newcommand{\bD}{\bm D}
\newcommand{\bd}{\bm d}
\newcommand{\br}{\bm r}
\newcommand{\bS}{\bm S}
\newcommand{\bg}{\bm g}
\newcommand{\wF}{\widetilde{F}}

\title{Electrodynamics of Amorphous Media at Low Temperatures}

\author{Vassiliy Lubchenko and Robert J. Silbey} 
\affiliation{Department of Chemistry, Massachusetts Institute of
Technology Cambridge, MA 02139}

\author{Peter G. Wolynes} \affiliation{Departments of Chemistry and
Biochemistry, and Physics, University of California, San Diego, CA
92093-0371}

\date{April 29, 2005}

\begin{abstract}

Amorphous solids exhibit intrinsic, local structural transitions, that
give rise to the well known quantum-mechanical two-level systems at
low temperatures.  We explain the microscopic origin of the electric
dipole moment of these two-level systems: The dipole emerges as a
result of polarization fluctuations between near degenerate local
configurations, which have nearly frozen in at the glass transition.
An estimate of the dipole's magnitude, based on the random first order
transition theory, is obtained and is found to be consistent with
experiment. The interaction between the dipoles is estimated and is
shown to contribute significantly to the Gr\"{u}neisen parameter
anomaly in low $T$ glasses. In completely amorphous media, the dipole
moments are expected to be modest in size despite their collective
origin.  In partially crystalline materials, however, very large
dipoles may arise, possibly explaining the findings of Bauer and
Kador, J. Chem. Phys. {\bf 118}, 9069 (2003).

\end{abstract}


\maketitle

\section{Introduction}

Glasses are frozen liquids and thus lack long-range order, yet the
differences in material properties between amorphous materials and
crystals are often rather subtle. Crystalline samples themselves are
rarely flawless and thus contain a number of imperfections such as
point defects, dislocations or grain boundaries of various
sorts. These tend to further mask the difference. The size of defects
in crystals ranges over many scales, while in glasses, the static
heterogeneity in the atomic arrangement appears comparable to the
molecular size itself.  Simple molecular glasses thus seem perfect
candidates for description as isotropic continuum, at long enough wave
lengths. For instance at cryogenic temperatures, when the de Broiglie
wave-length of a thermal phonon at $\sim$ 1K exceeds the lattice
spacing by three orders of magnitude or so, continuum theory would be
thought to hold to high accuracy. Yet surprisingly, there clearly
exist degrees of freedom numbering in great excess of the Debye
density of states, leading to extra heat capacity and phonon
scattering in all amorphous materials \cite{ZellerPohl}. Here we
examine the electrodynamics of these degrees of freedom.

Since Rayleigh scattering is too weak to account for the observed
magnitude of sound attenuation in glasses, internal resonances must be
involved, in the form of {\em anharmonic} structural rearrangments, in
order to explain the data. The well known, empirical two-level system
(TLS) theory presumes such resonances exist
\cite{AHV,Phillips}. Simply postulating a flat energy spectrum and a
frequency independent coupling to the phonons accounts for all the
gross features of the low $T$ anomalies (for reviews, see
\cite{LowTProp,HunklingerRaychaudhuri,Esquinazi}).  Direct microscopic
evidence of the two-level nature of such entities comes both from the
phonon echo experiments \cite{GG} and, relatively recently, from
single-molecule experiments at cryogenic temperatures (see
e.g. \cite{BTLBO}). At these temperatures, the TLS picture is
internally consistent in so far as the structural transitions (ST) can
be defined as {\em local}, and thus, tautologically, sufficiently {\em
weakly} interacting. One may therefore speak of a multilevel system at
the location of each transition whose behavior reduces to a TLS
behavior at low enough $T$. We may call this a tunneling center
(TC). In 1986, Freeman and Anderson \cite{FreemanAnderson} showed that
the magnitude of the TLS density of states is apparently correlated
with the phonon coupling. This results in a universality of the ratio
of the phonon mean free path $l_\smfp$ to its wave-length $\lambda$:
$l_\smfp/\lambda \sim 150$, for all insulating glasses at $T \lesssim$
1K. This universality seems hardly coincidental \cite{YuLeggett},
however understanding the origin of the universality requires a
microscopic picture of molecular motions in glasses. (The large size
of the factor $\sim$150, too, was a puzzle \cite{YuLeggett}.)

The Random First Order Transition (RFOT) Theory of the glass
transition \cite{KTW, MCT, MCT1, XW, LW_soft} provides an appropriate
microscopic picture of the motions in glass. Most commercial and
laboratory glasses are made by quenching supercooled melts. In the
deeply supercooled regime, most liquid motions are activated
transitions between distinct aperiodic states of comparable energy,
during which the current structural arrangement in a {\em local}
region is replaced by another, quite different arrangement that
nevertheless fits its environment. The size of the reconfigurable
region, $\xi$, grows with decreasing temperature, and reaches about
5-6 molecular units across by $T_g$, i.e. the glass transition
temperature corresponding to the 1 hour time scale. This size is
predicted to be universal, within logarithmic accuracy, for {\em all}
substances. At any point in time, above $T_g$, the liquid can be
thought of as a mosaic of such cooperative regions \cite{XW}, most of
nearly the same size, but otherwise with distributed barrier heights
and transition energies. Upon freezing, a particular mosaic pattern
sets in and undergoes relatively slow changes, called aging. The aging
speed depends on the quench depth \cite{LW_aging}. A sufficient
fraction of the structural transitions have small enough energy change
and barriers so that when they occur, they can account for the
cryogenic anomalies, some of which were mentioned above: the density
of states and the universality of phonon scattering \cite{LW}, the
Boson Peak \cite{LW_BP}, but also the anomalous Gr\"{u}neisen
parameter, the so called ``fast'' TLS systems and more
\cite{LW_thesis,LW_RMP}.  According to the RFOT theory, the
universality of the $l_\smfp/\lambda$ ratio directly follows from the
universal cooperative region size $(\xi/a)^3 \sim 200$ at the glass
transition temperature $T_g$, where $a$ is the molecular length
scale. During a structural transition, a relatively large, $\sim 200$,
compact set of small units moves in a stage-wise fashion. This
corresponds to the motion of the domain wall, which separates the two
alternative arrangements, through the compact region.  At cryogenic
temperatures, these motions occur by tunneling. Despite their
collective nature, such tunneling events are possible because of the
enormous multiplicity of alternative structural states and of
low-barrier paths connecting pairs of states: an amorphous sample
actually resides in a high energy density state, well above its lowest
energy, perfect crystalline state. Consistent with the facility of
tunneling is the smallness of individual atomic displacements during
each transition. Their amplitude is roughly equal to the Lindemann
length $d_L$. This length is typically one tenth of the characteristic
lattice spacing $a$ and is nearly the same for all substances. The
precise identity of the ``molecular unit'', or ``bead'' depends on the
specific substance, but usually corresponds to a few atoms. (See
\cite{LW_soft} for a detailed discussion.)

In this article, we use the microscopic picture of the two-level
systems provided by the RFOT theory to estimate the coupling of the
transitions to external electric fields. Clearly, such a coupling must
be present because individual molecular bonds, that possess electric
dipoles, rotate during transitions.  Since these couplings directly
enter into spectral hole-burning experiments \cite{Maier} and can also
be directly probed in single molecule experiments \cite{BTLBO}, it is
important to know how much collective excitations interact
electrodynamically with probes and external electric fields.

\section{Interaction of a Single Tunneling Center with External Field}

\subsection{Many-Body Origin of the Transition-Induced Dipole Moment}
\label{Motivation}

To set the stage, let us briefly review the assumptions of the
traditional molecular models of dielectric response of insulating
media.  One often assigns an electric dipole value to an individual
molecule, or to a molecular bond connecting distinct atoms in a
condensed phase.  In a dilute liquid made up of polar molecules, the
medium polarizes in a field since the dipoles prefer to orient along
the field's direction at the cost of losing their rotational
freedom. Even without permanent electric dipoles, a dielectric
response occurs due to polarizability: An external field mixes in
higher energy molecular orbitals, which generally lack inversion
symmetry. Classically, this quantum mechanical response can be
imitated as two harmonically bound opposite charges that separate
after a field is turned on. In a polar substance, this polarizability
also changes the {\em length} or {\em orientation} of the permanent
dipoles.
  
\begin{figure}[htbp!]
\includegraphics[width=.95\columnwidth]{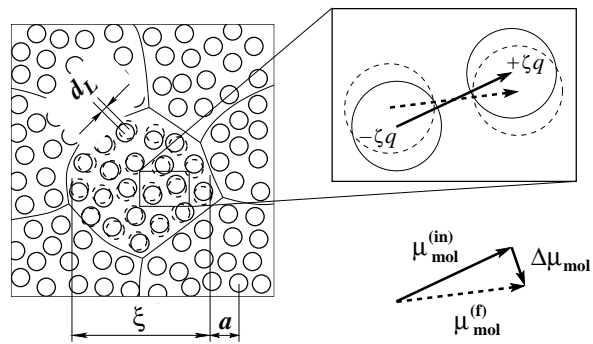}
 \caption{\label{domain_dipole} Shown on the left is a fragment of the
mosaic of cooperatively reconfiguring regions in the supercooled
liquid, with $a$ denoting the lattice spacing (more precisely ``bead''
spacing). $\xi$ is the cooperative region size; $d_L$ is the typical
bead displacement during a transition. (The shown magnitude of $\xi$
corresponds to a temperature near $T_g$ on 1 hour scale.) The two sets
of circles - solid and dashed ones - denote two alternative structural
states.  The expanded portion shows how rotation of a bond leads to
generating an elemental dipole change during a transition, where the
partial charges on the two beads are $\pm \zeta q$.}
\end{figure}

While in a fluid the dipoles can freely reorient, during a structural
transition (ST) in glass, the dipole is restrained: Each individual
bead within a tunneling center, or ``domain'', moves only about the
Lindemann length $d_L$, as illustrated in
Fig.\ref{domain_dipole}. Suppose, for the sake of argument, one can
break up the set of all the beads within the domain into distinct
pairs. During a transition, each such pair - and hence the
corresponding ``bond'' - rotates about $d_L/a \sim 0.1$ radian. The
amorphous lattice generally exhibits no symmetry. There is therefore,
typically, excess charge, however small, on each atom.  Assume the
effective individual charges remain the same during such a
transition. One can thus unambiguously assign a permanent, {\em
point-like} electric dipole to each ``bond'' introduced above.  As
schematically shown in Fig.\ref{domain_dipole}, a total dipole moment,
$ \bmu_\sT = \sum_i \Delta \bmu_i$, may be generated during a
structural transition, that would couple to an external electric field
$\bE$ with energy $- \bmu_\sT \bE$.

In molecular glasses, the bond dipoles are fairly easy to assign.
Generally, the assignment of point-like dipole moments to individual
bonds is, strictly speaking, non-unique. It would be rather difficult
in the case of a highly networked, covalently bonded substance, such
as amorphous silica, but as we shall see, other arguments for such
systems give similar results. These arguments use the measurable
piezoelectric properties of corresponding crystals to unambiguously
extract the coupling to fields.  In weakly bonded molecular glasses
held together by Van der Waals forces, the point like dipole view is
already a good approximation.

With this in mind, an order of magnitude estimate of the dipole moment
of a tunneling center can be made: The Coulomb charge on a bead does
not exceed a fraction $\zeta < 1$ of an elementary charge $q$, which
is close in magnitude to the electron charge $e$: $q \sim e$. An
individual electric dipole {\em change} is therefore $\Delta \mu \sim
\zeta \mu_\smol (d_L/a)$, where $(d_L/a)$ is the rotation angle, as
already discussed, and $\mu_\smol \equiv \zeta q a$ is the elemental
dipole magnitude associated with each bond. The number of pairs that
reorient in a structural transition is $N_\sd = (\xi/a)^3/2$. The
$\bmu_T$ distribution is, of course, centered at the origin. Since the
dipole forces are a small part of the energetics of the glass
transition, the dipole motions are expected to be only weakly
connected with each other.  Therefore the individual dipoles $\Delta
\bmu_i$ make up a ``random walk'' of $N_\sd$ steps (in 3D). As a
result, the generic value of dipole change for the transition is given
by the width of the total displacements during such a walk:
\begin{equation}
\mu_T \simeq \zeta (q a) [(\xi/a)^3/2]^{1/2} (d_L/a),
\label{mu_Tqual}
\end{equation} 
If the elementary dipole rotations are correlated, one may introduce
an additional factor - like the Kirkwood $g$ factor of liquid theory
\cite{Kirkwood}.  $g$ is typically of order 2. At first we might
imagine a great deal of variability for the quantity
$[(\xi/a)^3/2]^{1/2} (d_L/a)$ but in fact it is nearly universally (!)
equal to unity for all substances. $a$ is typically a couple of
angstroms, implying $q a$ corresponds to $\sim 10$ Debye. ($e \AA
\leftrightarrow$ 4.8 Debye.)  $\zeta$ is expected to be well less than
unity, with $0.1$ or less being a reasonable generic estimate. We thus
obtain that $\mu_T$ is of the order 1 Debye or less, consistent with
experiment. Note the magnitude of $\mu_T$ is rather modest - only of
order the size of a typical individual dipole moment - despite the
large number of particles constituting a domain. The relative
smallness of the two-level system dipole moment is due to the small
deflection angle $\sim 0.1$, and the apparent smallness of the partial
charge $\zeta q$. There are deep reasons for both: The former stems
from the particular magnitude of atomic displacements during a
transition: it is equal to the typical thermal displacement at the
mechanical stability edge, i.e. the Lindemann length \cite{XW}.  The
latter is probably related to the intrinsic difficulty of making
ionically bonded aperiodic structures, which imposes an upper bound on
the value of $\zeta$; this will be discussed in due time.
 
The sublinear scaling of the dipole moment with the domain volume
$\xi^3$, in Eq.(\ref{mu_Tqual}), is worth noting: The tunneling
transition dipole moment is not a bulk response, and its relation to
the material's average bulk dielectric properties, as encoded e.g. in
the substance's dielectric susceptibility $\epsilon(\omega)$ in the
fluid phase, is not immediately obvious. While it is ultimately the
deflections of the same elemental dipoles that both give rise to
$\epsilon(\omega)$ and the dipole moment, the causes of the
deflections differ. In contrast with the bulk polarizability, where
the deflection magnitude is proportional to the field, the transition
induced deflections are intrinsic and correspond to distinct local
structural states. (To illustrate this distinction further, we point
out that a transition can be induced by things other than an AC (!)
electromagnetic field - a thermal phonon for instance.) Above $T_g$,
the distinct structural states, that evolve into TLS at cryogenic
temperatures, are transient metastable structures that live typically
as long as the $\alpha$-relaxation time \cite{XW}. The transient
structures are (transiently) frozen-in elastic fluctuations.
Analogously, the intrinsic generated dipole moment may be thought of
as due to frozen-in electric fields - at $T_g$. (A formal connection
between generated electric field and mechanical stress is discussed in
the following Subsection.) We can use this line of thought to relate
the dipole moment to the bulk dielectric properties of the material
near $T_g$. Suppose the frozen-in field is along $z$ direction. The
corresponding dipole moment is related to this field via an
appropriate (frequency dependent) dielectric constant
$\epsilon_\sTLS(\omega)$, but also through the fluctuation-dissipation
theorem. A self-consistent closure of this relationship gives for a
spherical region of {\em volume} $\xi^3$ \cite{TitulaerDeutch}:
\begin{widetext}
\begin{equation}
- \frac{1}{k_B T_g} \int_0^\infty e^{- i \omega t} \frac{d}{dt} \la
 \mu_{\sT,z}(0) \mu_{\sT,z}(t) \ra \, dt =
 \frac{[\epsilon_\sTLS(\omega)-1][2 \epsilon_\sext(\omega)+1]}{4\pi[2
 \epsilon_\sext(\omega) + \epsilon_\sTLS(\omega)]} \xi^3,
\end{equation}
\end{widetext}
where $\epsilon_\sext(\omega)$ is the dielectric susceptibility
outside the domain. The dielectric constant inside,
$\epsilon_\sTLS(\omega)$, contains both the usual (high frequency)
polarizability of the material and the polarization due to the dipolar
displacements accompanying the transition. Even though the interior
and exterior of the domain are chemically identical, it is necessary
to regard the two $\epsilon$'s as distinct, since we know a {\em
transition} occurs within the volume $\xi^3$. A similar relation can
be written for the volume occupied by a single elemental dipole, with
$\epsilon_\sext(\omega) = \epsilon_\sTLS(\omega)$:
\begin{widetext}
\begin{equation}
- \frac{1}{k_B T_g} \left(\frac{d_L}{a}\right)^2 \int_0^\infty e^{- i
 \omega t} \frac{d}{dt} \la \mu_{i,z}(0) \mu_{i,z}(t) \ra \, dt =
 \frac{[\epsilon_\sTLS(\omega)-1][2 \epsilon_\sTLS(\omega)+1]}{4\pi[3
 \epsilon_\sTLS(\omega)]} (2 a^3),
\end{equation}
\end{widetext}
where the factor $(d_L/a)^2$ on the left reflects that the elemental
change in polarization, $\Delta \mu_i$, is related to the full
elemental dipole $\mu_i$ by the rotation angle $(d_L/a)$, see
Fig.\ref{domain_dipole}. The volume $2 a^3$ on the right corresponds
to the volume occupied by a pair of beads, as before. One gets, as a
result, a frequency dependent generalization of Eq.(\ref{mu_Tqual}):
\begin{widetext}
\begin{equation}
\la \mu_\sT^{(2)}(\omega) \ra = \la \mu_i^{(2)}(\omega) \ra
[(\xi/a)^3/2] \left(\frac{d_L}{a}\right)^2 \frac{[2
\epsilon_\sext(\omega)+1] 3 \epsilon_\sTLS(\omega)}{[2
\epsilon_\sTLS(\omega)+1][2 \epsilon_\sext(\omega) +
\epsilon_\sTLS(\omega)]},
\label{freq_dep}
\end{equation}
\end{widetext}
where the two-point correlation functions are the $t$ integrals above.
Note $\la \mu_\sT^{(2)}(0) \ra = \la \mu_\sT^2 \ra$ and $\la
\mu_i^{(2)}(0) \ra = \la \mu_i^2 \ra$. Finally, $\mu_i \simeq \zeta (q
a) \equiv \mu_\smol$, as before.

Note that two adjacent regions are statistically unlikely to undergo a
structural transition at the same time. The physically preferable
choice for the external dielectric susceptibility
$\epsilon_\sext(\omega)$ is therefore the high frequency, electronic
component of the full dielectric response, which we call
$\epsilon_\infty$. With this, Eq.(\ref{freq_dep}) becomes
\begin{widetext}
\begin{equation}
\la \mu_\sT^{(2)}(\omega) \ra = \la \mu_i^{(2)}(\omega) \ra
[(\xi/a)^3/2] \left(\frac{d_L}{a}\right)^2 \frac{[2 \epsilon_\infty+1]
3 \epsilon_\sTLS(\omega)}{[2 \epsilon_\sTLS(\omega)+1][2
\epsilon_\infty + \epsilon_\sTLS(\omega)]},
\label{freq_inf}
\end{equation}
\end{widetext}
Furthermore, the two-level systems that are active at low temperatures
correspond to the low barrier side of the barrier distribution. This
implies one should use the $\epsilon_\sTLS(\omega \rightarrow \infty)$
value with regard to the cryogenic phenomena and, therefore, no extra
frequency dependence appears in the coupling of the TLS to electric
field.  As a result, we find no significant reaction field correction
to our earlier argument and obtain
\begin{equation}
\la \mu_T^2 \ra \simeq \mu_\smol^2 [(\xi/a)^3/2] (d_L/a)^2,
\end{equation} 
c.f. Eq.(\ref{mu_Tqual}).

\subsection{Piezoelectric View}

To treat covalent network glasses, where the assignment of local
dipoles or changes is difficult, we now turn to a different way to
relate the dielectric response, within a domain, to frozen mechanical
stress. First consider a strictly periodic lattice that lacks parity
symmetry. Generally, the lattice strain induces internal electric
fields giving rise to piezoelectric behavior. In such a piezoeletric,
the energetics of the strain, in the lowest order, are described by
the free energy density (see e.g. \cite{LLcont})
\begin{equation}
\wF = \frac{1}{2} \lambda_{ik,lm} u_{ik} u_{lm} - \frac{1}{8 \pi}
\epsilon_{ik} E_i E_k + \beta_{i,kl} E_i u_{kl}.
\label{quadr1}
\end{equation}
Here $u_{ik}$ is the standard strain tensor \cite{LLelast},
$\lambda_{ik,lm}$ and $\epsilon_{ik}$ are the stiffness tensor and the
dielectric tensor respectively, and $\beta_{i,kl}$ is {\em a}
piezoelectric tensor. (Note, various sign conventions and free
energies have been used in the literature.) The double index summation
convention is implied throughout, with the exception of letters $x$,
$y$, and $z$, which will be obvious in the context.  Finally, the
elastic constant $\lambda$ is related to the material's mass density
$\rho$ and the speed of sound $c_s$:
\begin{equation}
\lambda \sim \rho c_s^2.
\label{stiffness}
\end{equation}

In the absence of external field: $\bD = - 4 \pi (\prtl \wF/\prtl \bE)
= 0$, - the internal electric field is simply proportional to the
strain itself:
\begin{equation}
E_i = 4 \pi \epsilon^{-1}_{ik} \beta_{k,lm} u_{lm}.
\label{E_spont}
\end{equation}
(Note, the $\beta$ tensor has the dimensions of electric field.)  If
expressed in terms of strain only, the free energy density reads, in
the absence of external electric field:
\begin{eqnarray}
\wF &=& \frac{1}{2} \lambda'_{ij,kl} u_{ij} u_{kl}, \label{quadr2} \\
\lambda'_{ij,kl} &\equiv& \lambda_{ij,kl} + 4 \pi \epsilon^{-1}_{mn}
\beta_{m,ij} \beta_{n,kl}. \label{lambda_pr}
\end{eqnarray}
The total, apparent stiffness, $\lambda'$, can be decomposed thereby
into a purely ``covalent'' and a ``coulomb'', i.e. electrostatic
component. The latter contribution is ordinarily quite small, owing to
the smallness of the charges induced by lattice distortions. Consider
$\alpha$-quartz, for example. Here, only $\beta_{x,xx} = 5.2 \cdot
10^4$ esu, and $\beta_{x,yz} = \beta_{x,zy} = (1/2) 1.2 \cdot 10^4$
esu, are non-zero \cite{Cady}. As a result, the ``coulomb''
contribution to the $(xx,xx)$ component of the apparent stiffness
tensor is only about one percent of the covalent counterpart:
$\lambda_{xx,xx} = 8.8 \cdot 10^{11}$ dyne/cm$^2$ and $(4
\pi/\epsilon_1^T) \beta_{x,xx}^2 = 7.4 \cdot 10^9$ dyne/cm$^2$.
(Here, we used $\epsilon^T_1 = 4.58$ \cite{Mason}.)  The relative size
of the $\lambda$ vs. $\beta$ magnitudes can be understood as follows:
$\lambda$ reflects the energy (density) of the elastic restoring
force.  It is essentially the second derivative of an individual
atomic potential. Interatomic bonding, be it coulombic or covalent in
character, is ultimately of electrostatic origin. One may therefore
associate $\lambda$ with the quantity $\frac{1}{a^3}
\frac{\prtl^2}{\prtl^2 r} (q^2/r)|_{r=a} = (q/a^2)^2$, where $q$ is
the effective charge giving rise to the bond, and the $1/a^3$ factor
in front provides for energy density. (The total {\em first}
derivative of the (full quantum) potential energy is zero, of course.)
By virtue of being an electric field, $\beta$ roughly corresponds to
the quantity $q'/a^2$, where $q'$ would be the partial charge
introduced in the previous subsecton. The ratio $q'/q$ corresponds,
within the present framework, to the earlier introduced quantity
$\zeta$.  It follows that in $\alpha$-quartz, the partial atomic
charge is indeed about one tenth elementary charge, since the quantity
\begin{equation}
\zeta^2 = (q'/q)^2 \simeq \beta^2/\lambda 
\label{zeta2}
\end{equation}
is approximately equal to 1/100 in silica, as we just saw.

Now suppose for a moment that a relation similar to Eq.(\ref{E_spont})
exists between the bead displacements within a domain and the internal
electric field changes generated during a transition. (We stress, in
an amorphous sample such generated field changes are zero, upon
spatial average, but here we refer to {\em local} fields at a
particular, generally non-centrosymmetric site.) Since $d \wF = - \bD
d \bE/4\pi + \ldots$, the free energy change in the presence of a
(small) external field $\bD_\sext$ during the transition is given by
$\int_V dV \, \Delta \wF = - \bD_\sext \int_V dV \, \Delta \bE/4\pi$,
in the lowest order in $\bD_\sext$. (Here we have volume-integrated
over the reconfiguring domain that correspons with the two-level
dynamics; $\bD_\sext$ obviously varies sufficiently slowly within the
domain for realistic frequencies of light, and can be taken out of the
integral.)  The relation of the field $\bD_\sext$ to the external
field proper depends, of course, on the experiment's geometry.  We
will use an electric field $\bE_\sext (\br) = \epsilon^{-1}
\bD_\sext(\br)$ where $\epsilon$ is the average {\em bulk} dielectric
susceptibility (which, of course, is uniform and isotropic in an
amorphous material). The coupling to this field $\bE_\sext$ is
consequently given by:
\begin{equation}
\bmu_T = \frac{\epsilon}{4 \pi} \int_V dV \, \Delta \bE(\br),
\label{muT_formal1}
\end{equation}
i.e. the generated internal field difference during the transition,
integrated over the domain.  Eq.(\ref{muT_formal1}), among other
things, demonstrates that one may indeed unambiguously assign a
collection of point-like dipoles to the bead set within a TC, namely
by virtue of the relation $\bE(\br) = 4 \pi c_1 \bmu(\br) n(\br)$,
where $n(\br)$ is the (coordinate-dependent) dipole concentration and
the constant $c_1 \sim 10^0$ should be chosen depending on a specific
way to incorporate the already mentioned cavity effects. In what
follows, we will outline the microscopic picture of interaction of a
transition with elastic strain, which will naturally lead us to the
formula above and the ability to estimate the electric moment via
material's piezoelectric properties.

Since individual displacements $\bd_i$ during tunneling transitions
are only one-tenth of the lattice spacing, one can indeed describe the
corresponding additional elastic energy variations, due to the
presence of a phonon, by a quadratic form of the type from
Eq.(\ref{quadr1}) or (\ref{quadr2}).  Further, here one computes {\em
relative} displacements, not the absolute atomic coordinates which are
generally difficult to calculate.  Define $\phi_{ik}$ as the strain
tensor due to a (long-wave) lattice distortion of a stable lattice. In
addition, define $d_{ik}$ as the ``strain'' tensor corresponding to
the set of the tunneling displacements $\{\bd_i\}$.  The full elastic
energy within the domain, given a particular domain boundary
configuration (call it $\Omega_b$) can be written as:
\begin{equation}
\wF = \frac{1}{2}\lambda'_{ij, kl}
[(\phi_{ij}+d_{ij})(\phi_{kl}+d_{kl}) - d_{ij} d_{kl}] + {\cal
H}(\{d_{ij}\}, \Omega_b),
\end{equation}
where the energy functional ${\cal H}(\{d_{ij}\}, \Omega_b)$ includes
all the non-linear, many-body interactions giving rise to the
existence of the many metastable structural minima within the
domain. The construction of a library of the states corresponding to
these minima was described in Ref.\cite{LW_aging}. The full
multiplicity of the local states reveals itself directly in
calorimetric measurements above the glass transition. Here, we are
only concerned with the two lowest energy states of a region of the
otherwise undisturbed lattice. The two correspond to the two lowest
minima of ${\cal H}(\{d_{ij}\}, \Omega_b)$. The size of the domain is,
as we have seen, chosen so that one is guaranteed to have at least one
alternative structural state of nearly the same energy, and was found
to be only slightly larger that the cooperative region size at $T_g$
\cite{LW}. Note, by construction, the boundary state $\Omega_b$ is
independent of the phonon field $\phi_{ik}$. The effect of an external
{\em mechanical} stress on the internal displacements within a local,
compact region is passed on through the boundary, and so the
interaction of the region with the stress can be expressed through a
displacement integral over the region surface \cite{LLelast}. The
cross $\lambda_{ij,kl} \phi_{ij} d_{kl}$ term in Eq.(\ref{quadr2})
gives the amount by which the energy of a tunneling transition is
modified by the presence of a phonon. This therefore gives the
TLS-phonon coupling. The latter coupling was estimated in this way in
Ref.\cite{LW,LW_RMP}. A direct computation of the
$\phi_{ij}$-field-induced change of the transition energy gives:
$\Delta E(\phi_{ij}) = \lambda'_{ij,kl} \phi_{ij} \int_V dV \,
d_{kl}$, where we have integrated over the domain volume and taken
advantage of the elastic strain being $\phi_{ij}$ nearly constant
throughout the domain ($\lambda'$ in the latter equation is the
domain-averaged value of the atomic force-constant.) The coefficient
at the $\phi_{ij}$ gives a (tensorial) coupling of a transition to
strain according to:
\begin{equation}
{\cal H}_{\sTLS, \sph} = g_{ij} \phi_{ij} \sigma_z,
\label{H_int}
\end{equation}
where $\sigma_z = \pm 1$ is the usual Pauli matrix, and
\begin{equation}
g_{ij} = \frac{1}{2} \lambda'_{ij, kl} \int_V dV \, d_{kl}.
\end{equation}
The volume integral above indeed reduces to a surface integral of the
tunneling displacements. Consider for example, the term $\int_V dV \,
d_{xy} = \int_S (d_x dx + d_y dy) dz$ etc. The coupling to the {\em
longitudinal} phonons has the most vivid form, since $d_{ii}$ is the
divergence of a vector field, i.e. $\bd$. One gets $g_{ii} \simeq
\lambda \int_S \bd \, d\bS$. The atomic displacements at $T_g$ are
typically near the Lindemann length: $d \simeq d_L$, - but can also be
expressed in terms of the elastic constants, since the amount of
elastic energy contained in a unit cell is determined by the
temperature itself: $a^3 \lambda (d/a)^2 \sim T_g$. (This is by virtue
of the fluctuation-dissipation theorem.)  Estimating the surface
integral \cite{LW, LW_RMP} introduces additional numerical factors and
gives, within a factor of two or so:
\begin{equation}
g \simeq \sqrt{T_g \rho c_s^2 a^3},
\label{g}
\end{equation}
where we have used Eq.(\ref{stiffness}). The result above is easy to
rationalize on general grounds: Any atomic motions in a dense liquid
(at $T_g$ and otherwise) are either a vibration or an anharmonic
motion that is part of a structural transition. The two excitations
must coexist and thus be marginally stable against each other, in
order for both to be present. Such a marginal stability criterion
gives $\la \sigma_z g_{ij} \phi_{ij} \ra \simeq \la \lambda_{ij, kl}
\phi_{ij} \phi_{kl} \ra$, as follows from optimizing Eq.(\ref{H_int})
together with the elastic energy $\frac{1}{2} \lambda_{ij,kl}
\phi_{ij} \phi_{kl}$ with respect to $\phi_{ij}$, multiplying by
$\phi_{ij}$ and thermally averaging (at $T=T_g$). Owing to $\la
\sigma_z \phi_{ij} \ra \simeq \la |\phi_{ij}| \ra$, Eq.(\ref{g})
follows.

In the same way that the elastic fluctuations interact with the atomic
displacements, polarization waves and external electric sources will
interact with the internal electric fields generated in the domain
during a transition. As we have seen, the mechanical response
characteristics of a transition arise in response to stress
fluctuations at $T_g$. Analogously we can say, the electric moments of
the two-level system arise in response to the electric field
fluctuations at $T_g$. In full analogy with the
fluctuation-dissipation theorem context, the TLS dipole moment will
interact with an external field source just as it did with the
internal electric fields at the moment of freezing. We may thus
compute the dipole moment by substituting the generated electric field
from Eq.(\ref{E_spont}) into Eq.(\ref{muT_formal1}), bearing in mind
that the local lattice and the corresponding tensors are no longer
subject to any particular symmetry. Still, since the lattice locally
resembles a crystalline lattice, one may choose coordinates, again
locally, in such a way that the the $\beta$ tensor is maximally close
to a crystalline one. (Clearly, the $m$, $n$ sum in
Eq.(\ref{lambda_pr}) is independent of the coordinate choice.)
Therefore the latter sum will give a comparable result to that for a
crystal.  Summing over the displacement tensor $d_{ij}$ (contracted
with the $\beta$ tensor) is quite analogous to the argument leading to
Eq.(\ref{mu_Tqual}). As a result one obtains the following qualitative
estimate:
\begin{equation}
\mu_T \simeq (\bar{\beta} a^3) [(\xi/a)^3/2]^{1/2} (d_L/a),
\label{mu_Tpiezo}
\end{equation} 
where $\bar{\beta}$ is the local value of the piezoelectric
coupling. (We remind the reader that glasses are on average isotropic
and thus can not have bulk piezoelectric properties. It is only in the
frozen state that parity is locally broken.) The simple relation
\begin{equation}
\zeta q a \simeq \bar{\beta} a^3
\label{connection}
\end{equation}
establishes the connection between the ``molecular dipole'' view of
the previous Subsection and the ``piezoelectric'' analysis in this
Subsection, c.f. Eq.(\ref{zeta2}). Using the quartz parameters above,
one obtains that $\mu_T$ is, again, of the order Debye. Note that
formula (\ref{mu_Tpiezo}) uses quantities that can be measured
independently for substances which have a crystalline counterpart.
The bead size $a$ can be determined from the fusion entropy
\cite{LW_soft}. The piezoelectric constants are measurable too.  The
two views - one based on molecular moments, the other on local
piezoelectricity - are somewhat distinct but are highly overlapping:
Since the local hybridization pattern on individual atoms is
intrinsically asymmetric in amorphous lattices, partial atomic
charges, however small, are always expected to be present in glasses,
giving rise to both local permanent dipoles {\em and} local
piezoelectricity. Mixing in a dipolar species would enhance both
effects.  According to Ref.\cite{Schickfus_OH}, the two-level
systems's dipole magnitude is very correlated - nearly proportional -
to the OH$^-$ ion concentration, in amorphous silica with OH$^-$
impurities. Yet extrapolation to small ion concentrations shows TLS in
silica exhibit an {\em intrinsic} dipole moment, as was later
confirmed by an electric dipole echo study \cite{Golding_dipoleecho}.

\subsection{Electrodynamics and Electroacoustics: Connection with
Experiment}

In order to discuss experiments of two-level systems in glasses,
involving external fields, let us first recapitulate a few aspects of
the traditional phenomenological description, along with the
microscopic explanation.  The Hamiltonian of an isolated two-level
system, as usually written in the low $T$ glass context, is
\begin{equation}
{\cal H}_\sTLS = \frac{1}{2} \epsilon \sigma_z + \frac{1}{2} \Delta
\sigma_x + g_{ij} \phi_{ij} \sigma_z,
\label{H_TLS}
\end{equation}
where $\epsilon$ is the transition energy, $\Delta$ is the tunneling
matrix element. (The phonon part of the full Hamiltonian is given in
Eq.(\ref{quadr2}).) According to Refs.\cite{LW, LW_BP}, the TLS that
are thermally active at cryogenic temperatures have their splitting
$\epsilon$ distributed according to a simple Boltzmann-like law
$n(\epsilon) = \frac{1}{T_g} e^{-\epsilon/T_g}$. This roughly defines
the density of states of the tunneling transitions. The density of
states, as seen by calorimetry, is time dependent, because the
tunneling matrix elements $\Delta$ are widely distributed. According
to the semi-classical analysis in Refs.\cite{LW, LW_BP}, the
distribution is $P(\Delta) \propto 1/\Delta^{1+c}$, where $c \ll 1$ is
a small constant ($c \propto \hbar \omega_D/k_B T_g$). This
distribution is close to, but not precisely the same as the inverse
distribution $\propto 1/\Delta$ postulated by the phenomenological TLS
model. Including quantum effects reveals that the low splitting
two-level systems, i.e. those with $\epsilon \ll \Delta$, are special
in the following sense. In such regions, the excess strain energy of
the glass is concentrated in the domain wall itself, while the barrier
separating the two alternative structural states is not high enough to
keep the domain in any of the classical structural states as defined
in terms of the classical atomic coordinates. Such TLS, with depinned
domain walls, give rise to an extra piece in the combined $P(\epsilon,
\Delta)$ distribution \cite{LW_RMP}. They correspond to the so called
``fast'' two-level systems, introduced early on phenomenologically
\cite{BlackHalperin} in order to rationalize certain quantitative
shortcomings of the original TLS model. The quantum depinning of the
domain wall has been also called ``quantum mixing'' \cite{LW_RMP}.
Finally, we note the Hamiltonian in Eq.(\ref{H_TLS}) leads to rich
relaxational behavior, due to both interaction with phonons and the
phonon-mediated interaction with other TLS's. This has been discussed
in detail previously \cite{relax}, \cite{HunklingerRaychaudhuri},
\cite{Phillips}.

The effects of the interaction of the dipole moment with a {\em
static} electric field are actually quite difficult to observe under
routine laboratory conditions (see e.g. \cite{Stephens_dipole}). The
upper limit for the field is given by the dielectric breakdown value
and is generically $10^6$ V/cm. For the typical dipole moment of 0.5
Debye, this implies an interaction energy of only 10$^{-3}$ eV $\sim$
10 Kelvin. The constant field strength normally employed is actually
an order of magnitude weaker, or less. In a field $\bE_\sext$, the
transition energy is modified according to $\epsilon \rightarrow
(\epsilon - \bmu_\sT \bE)$. Typically, $|\bmu_\sT \bE| < 1$K. This is
clearly inferior to the characteristic energy scale of the TLS
spectrum, namely the glass transition temperature $T_g$ \cite{LW}. The
effect of a constant field thus turns out to be very generic because
of the intrinsic flatness of the energy distribution: The angular part
of $\bmu_\sT \bE_\sext$ is, obviously, uniformly distributed
resulting, again, in a flat distribution of the field-modified
transition energy. In order to discern such a small energy variation,
a {\em resonance} technique must be employed. Just such an experiment
was performed by Maier at el. \cite{Maier}, who took advantage of the
possibility to burn very narrow holes - only a few MHz - in the
chromophore's inhomogeneous spectrum. These authors turn on the field
immediately after burning the hole and observe the (time-dependent)
hole broadening, whose overall magnitude depends quadratically on the
field strength. Maier at el. report the value of $\mu_T = 0.4$ D for a
PMMA matrix.

It follows from our theory that light and sound couple to the
two-level system transitions in a very similar way, save the dipole
character of the TC-photon interaction distinct from the tensorial
coupling of the transitions to the phonons.  Indeed, the temperature
dependence of the speed of light in vitreous silica, as obtained early
on in by Schickfus at el. \cite{Schickfus}, nearly coincides with the
corresponding ultrasonic data \cite{HunklingerRaychaudhuri}. This
strongly suggested, at the time, that both the electromagnetic and
acoustic anomalies had the same origin. That the coincidence is not
purely circumstantial was shown soon afterwords in a number of elegant
electro-acoustic experiments: Increasing the AC electromagnetic field
leads to saturation of the structural transitions and a decrease in
ultrasonic attenuation \cite{Laermans, Doussineau}. In addition,
exposure to the AC field affects the {\em acoustic} impedance of a
glass \cite{Doussineau}. Again, the sufficient sensitivity of these
experiments is due to the interaction with the AC field being
resonant. Finally we mention yet another venue in investigating the
TLS coupling to electric fields, namely the electric dipolar echo, see
e.g. \cite{Bernard_dipoleecho, Golding_dipoleecho}.

The dipole moment magnitudes, reported in all these experiments on the
respective substances, are all of the order 1 Debye, although more
recent measurements seem to be converging on a fraction of a
Debye. Unfortunately, the extracted dipole values do not completely
agree between different experiments. So, for example, Kharlamov at
el. \cite{Kharlamov_dipole} give relatively low values of 0.2 D and
0.1 D for PMMA and PS respectively, based on their field induced
spectral diffusion data.  The degree of the quantitative discrepancy
is, of course, subject to the detailed assumptions on the distribution
of the individual TLS parameters, various angular averagings etc.

\section{Dipole-Dipole Interaction}

The idea of local structural transitions is internally consistent in
that the transitions are indeed distinct, weakly interacting entities.
This is easy to understand by considering the moment of vitrification,
when a particular pattern of mobile regions sets in: A transition will
be found locally, upon freezing, if at $T_g$ it was of marginal
stability with respect to external mechanical perturbation, as
delivered by stress waves to the given local region. It does not
matter, of course, whether the source of these waves is thermal
elastic fluctuations or the other structural transitions. Now, upon
having estimated the energy spectrum of the TLS and their coupling to
the phonons \cite{LW}, one may check the magnitude of the resultant
TLS-TLS interaction, mediated by the acoustic waves in the frozen
lattice. Such interaction self-consistently turns out to be small
\cite{LW_RMP}. We are aware of several observable consequence of the
interaction. For one thing, this interaction tends to quench the
spontaneous echo generation \cite{BlackHalperin, GG}, by virtue of
dephasing each TLS's motion when the TLS precesses about its local
field. Another, remarkable effect from the interaction is that it
gives rise to a negative thermal expansion coefficient in some
glasses, at low enough temperatures \cite{LW_RMP}: The fluctuating
entities in the lattice attract in the Van der Waals fashion, via
exchanging phonons. This attraction, counterbalanced by the materials
stiffness, acts to partially contract the sample. The number of
thermally excited transitions increases with the temperature and
thereby enhances the degree of contraction.  The effect is small,
about $10^{-6}$, but nevertheless observable. The often employed
dimensionless parameter characterizing lattice non-linearity - the
Gr\"{u}neisen parameter - is usually positive and of the order unity
in crystals. It was found to be large and negative in many glasses at
cryogenic temperatures, see e.g. \cite{Ackerman}.

The direct phonon mediated TLS-TLS interaction goes as $1/r^3$, just
like the usual dipole-dipole interaction, but is somewhat complicated
by the tensorial form of the coupling (see e.g. \cite{BlackHalperin}
for a discussion). If, however, the transverse and longitudinal speeds
of sound were equal, the interaction would be strictly dipole-dipole.
It is therefore often convenient to assume the elastic interaction is
indeed purely dipole-dipole resulting in a small {\em quantitative}
error.  With this simplification, a ``scalar'' phononic Hamiltonian
can be used: $\frac{1}{2} \lambda_{ij, kl} \phi_{ij} \phi_{kl}
\rightarrow \frac{1}{2} \rho c_s^2 (\nabla \phi)^2$, where $\phi$ is a
scalar displacement field polarized in a single direction. The
coupling will become $g_{ij} \phi_{ij} \sigma_z \rightarrow (\bg
\nabla \phi) \sigma_z$. (The tensorial character of the interaction
may actually be important in the Gr\"{u}neisen parameter context, see
\cite{LW_RMP} and below). The presence of an electric component to
each transition dipole moment clearly leads to another contribution to
the total interaction.  Since a detailed discussion of the interaction
effects has already been given elsewhere \cite{LW_RMP}, here we simply
estimate the strength of the electric dipole-dipole coupling relative
to the purely elastic counterpart and the rest follows in a
straightforward fashion.

The elastic dipole-dipole interaction is given by a simple formula,
see e.g. \cite{LW_RMP}, \cite{YuLeggett}:
\begin{equation}
{\cal H}_{\selast} \simeq \frac{g^2}{\rho c_s^2} \frac{1}{r^3}
\simeq T_g \left( \frac{a}{r} \right)^3,
\label{elast}
\end{equation}
where, for the sake of clarity, we eschew some numerical constants
(these could be found in Ref.\cite{LW_RMP}) and have used
Eq.(\ref{g}). The electric dipole-dipole interaction, by
Eqs.(\ref{zeta2}) and (\ref{connection}), is, on the other hand:
\begin{equation}
{\cal H}_{\select} \simeq \zeta^2 \rho c_s^2 a^3 \left( \frac{a}{r}
\right)^3.
\label{elect}
\end{equation}
What is the relative value of the two interactions? A useful rule of
thumb is that $g \simeq \sqrt{T_g \rho c_s^2 a^3}$ is of the order eV
for all substances.  In silica, for instance, $\rho c_s^2 a^3$ is
several eV, the Rydberg scale being a convenient (and physically
justified) landmark. The $T_g$ of silica is 1500 K, i.e. slightly
larger than 0.1 eV. We therefore make an interesting observation that
the electric dipole-dipole interaction {\em can} be comparable in
magnitude to the elastic counterpart for polar enough substances.
This is despite the relatively weak contribution (1\% or less) of the
coulombic forces to the apparent mechanical stiffness. We will
speculate on the physical significance of this observation in the
final Section of the article, while for now, we limit ourselves to a
formal notion: The elastic dipole-dipole interaction is disadvantaged,
compared to the electric counterpart, due to the large $\rho c_s^2$
term in the denominator of Eq.(\ref{elast}): Phonons are not true
gauge particles.

Leggett has emphasized \cite{Leggett} that the dimensionless
Gr\"{u}neisen parameter varies ``wildly'' between different amorphous
substances, in contrast with the nearly universal $l_\smfp/\lambda$
ratio. Lubchenko and Wolynes have argued \cite{LW_RMP}, this stems
from the Van der Waals attraction between the tunneling centers, which
is strongly enhanced by ``Boson Peak'' excitations. The total
attractive interaction consists of several contributions, is
temperature dependent, and is expressed in terms of various
combinations of the temperature, $T_g$ and the Debye
temperature. While it may be argued that there is an intrinsic {\em
upper bound} on the value of $\zeta$ (see below), there otherwise
seems to be little intrinsic connection between the polar and
structural characteristics of glasses, in general. As a result, the
electrostatic interaction from Eq.(\ref{elect}) is expected also to
contribute to the non-universality of the Gr\"{u}neisen parameter.
This notion is consistent with the sensitivity of the Gr\"{u}neisen
parameter in silica to the concentration of polar impurities
\cite{Ackerman}. Further, the magnitude of the negative thermal
expansivity is indeed larger in more polar mixtures according to
Ref.\cite{Ackerman}.

{\em Symmetry vs. Transient Piezoelectricity.} We have so far focused
on the dipole moment of one of the structural states {\em relative} to
the other, namely $\sum_i \Delta \bmu_i = \mu_\sT$, since it is what
determines the coupling of the transition to external field. We next
inquire whether there is a correlation between the degree of polarity
of a state with its absolute energy. For example, suppose for a moment
that the lower energy state is completely non-polar, so that the
two-level system dipole moment is completely due to the excited state.
Such a situation could be exploited experimentally: One could
supercool a liquid just enough so that it does not crystallize readily
(the way glassblowers do), then expose the sample to strong electric
field for a sufficient while, and then quench the sample below its
$T_g$. After that, remove the field.  Clearly, the number of dipole
moments along and opposite to the field direction will differ. In
other words, a number of dipoles will be actually lined up in a
preferred direction, leading to a (weak) ferro-electric order. As a
consequence of this, removal of the field in the procedure above
should lead to sample's contraction (which is the sample's way of
minimizing the ferroelectric energy.) Some residual polarization will
appear as well.  If the field is removed sufficiently fast, the sample
will also heat up some. (This, in a sense, represents an antithesis to
adiabatic demagnetization.)  Such polarization will not take place if
the degree of polarity is uncorrelated with the energy of a structural
state, at least within the relevant energy range.

While we can not, at present, rule out {\it \`{a} priori} the
``pyroelectric'' scenario above, it seems rather unlikely for the
following reason. An amorphous sample, at least at long enough times,
should have inversion symmetry, on average. With this symmetry
present, no piezoelectricity, let alone pyroelectricity, can take
place \cite{LLcont}.  The question therefore is whether an external
field breaks the inversion symmetry (locally) without inducing
crystallization. Perhaps it will, on short enough time scales, before
considerable aging takes place. An experimental study would settle
this issue. At any rate, even if present, ferroelectric order would
not affect the thermal expansion properties at low temperatures
(c.f. the discussion of the Gr\"{u}neisen parameter in
\cite{LW_RMP}). This attraction mechanism is temperature independent
and simply contributes to the effective molecular field at each TLS
site.

\section{Closing Remarks}

In this article, we have outlined the microscopic origin of the
coupling of the intrinsic structural transitions in amorphous solids
to electric fields. The coupling stems from rotation of the molecular
bonds, within the region encompassing the transition, which generates
a net electric dipole. The molecular constituents of a glass, even if
intrinsically non-polar, are strained due to disorder. Therefore small
partial charges on each atom are expected leading to the presence of
electric dipole moments associated with individual bonds.

A local structural transition occurs by moving a domain wall through
the region. The domain wall is a mechanically strained region
separating alternative structural states. Such strained regions are
frozen-in thermal fluctuations of the lattice. (Above $T_g$, the
life-time of such as frozen-in structure would not exceed the typical
$\alpha$-relaxation time in the liquid.) In an analogous fashion, the
dipole moment can be thought of as frozen-in polarization
fluctuations, or as the local piezoelectric response to local strains.

The glass transition is driven by steric, i.e. mechanical
interactions, not primarily by electrical ones, which is reflected in
the smallness of the partial charge, given in the theory by the
dimensionless quantity $\zeta$.  So, for example, the elastic modulus
can be represented as $\rho c_s^2 (1 + \zeta^2)$, where $\zeta^2
\lesssim 0.01$ is the contribution of the ionic forces to the overall
material stiffness. (Here, we discriminate between ionic (or hydrogen)
bonds, as in NaCl or H$_2$O, and covalent bonds, as in diamond.) In
spite of its apparent small contribution to the material's structural
integrity, the coulomb component could actually be comparable to the
elastic component of the interaction between distinct {\em
transitions}. Note such interactions are unimportant as far as the
identity of each transition is concerned, because the corresponding
regions are statistically far apart \cite{LW_RMP}. Nevertheless, this
weak interaction may be viewed, loosely, as the successful attempt of
the system to have avoided entropically costly local ferroelastic
order. With this in mind, and the relatively strong coulomb
interaction of structural fluctuations, the smallness of the partial
charges in many glasses may simply reflect the fact that purely ionic
compounds, such as sodium chloride or water, do not vitrify readily
but instead, form low density lattices.  (Note, ice does amorphize
under high pressure, see e.g. \cite{Strassle}.)  Conversely, this
imposes an upper bound on the $\zeta^2$ value.

Consider now Eq.(\ref{lambda_pr}). In order for the lattice to be
stable, the $\lambda'$ matrix (with an $ij$ pair considered a single
index) should be positive definite. This means, if a particular
$\lambda$ is negative, the corresponding $\beta$ could be, in
principle, quite large. This would imply large induced dipole moments.
As just argued, the coulomb component is likely to be small in a
glass, because the latter is rather homogeneous. Defects in crystals,
however, can be more extended and can be highly anisotropic. In this
regard we wish to mention the single chromophore studies of spectral
drift in a Shpolskii matrix, by Bauer and Kador \cite{BauerKador}.
These authors have seen a transition (presumably due to a structural
defect) that generated a remarkably large dipole moment of 8
Debye. Note, the Shpolskii matrix is polycrystalline, not strictly
amorphous.

Finally we remark there is more to electrodynamics of amorphous solids
than what could realistically be discussed in this communication. Here
we have analysed only the case of good electric insulators. Distinct,
interesting phenomena take place in semiconductors and metallic
glasses. As an example, let us mention an old experiment of Claytor
and Sladek \cite{ClaytorSladek}, who found enhanced ultrasonic
attenuation in As$_2$S$_3$ glass upon removal of electric field. This
extra attenuation was greatly reduced by infrared radiation, which
suggested that there is, we quote, ``atomic relaxation accompanying
electronic transition in gap states where injected carriers have been
trapped''. We leave this for future work.

{\em Acknowledgments:} P.G.W. was supported by the NSF grant CHE
0317017. V.L. and R.J.S. gratefully acknowledge the NSF grant CHE
0306287 and the Donors of the Petroleum Research Fund of the American
Chemical Society. We are happy to dedicate this bagatelle to Bob
Harris, whose elegant contributions are many, but more than scan the
range from the electrodynamical properties of matter, parity, and
tunneling systems, all of which enter the question discussed here.

\bibliography{/home/vas/tex/lowT/lowT}

\end{document}